# Lost work and exergy: two views of the same concept

Joaquim Anacleto

Departamento de Física da Escola de Ciências e Tecnologia da Universidade de Trás-os-Montes e Alto Douro, Quinta de Prados, 5000-801 Vila Real, Portugal

IFIMUP-IN e Departamento de Física e Astronomia, Faculdade de Ciências, Universidade do Porto, R. do Campo Alegre s/n, 4169-007 Porto, Portugal

e-mail: anacleto@utad.pt

ORCID: orcid.org/0000-0002-0299-0146

## Abstract

We present the fundamental equation for a system and for a process, and by considering irreversibility within the system, we show that the *lost work* concept emerges naturally from the formalism. We then argue that if irreversibility is considered within the surroundings the lost work becomes what is known as *exergy*. Therefore, lost work and exergy are two views of the same concept, which in turn integrates a broader and more fundamental concept: *entropy generation*. It is our opinion that the clarification of the meanings of lost work and exergy, as well as the discussion that leads to an understanding of their differences and similarities, has not received the attention in the literature that it deserves. This paper fills that gap, and it is hoped that the discussion of these two concepts here will be useful for both students and teachers.

## 1. The fundamental equation

Many pedagogical difficulties in thermodynamics are tacitly accepted because we are so used to them. In particular, the abundance of terms, some of which are often used as synonyms for the same ideas, leads to a conceptual structure that is difficult to understand.

In this study, we focus on the concepts of lost work and exergy. The former is more common in physics, while the latter appears mainly in engineering, a discipline in which thermodynamics has many important applications. In the literature, the differences and similarities between these two concepts, as well as their definitions, have not been made clear. To overcome ambiguities, points of confusion or even errors regarding these concepts, we will use logical-mathematical reasoning and its intrinsic critical spirit, rather than accepting difficulties as subtleties while developing intuition-based representations.

Thermodynamics deals with macroscopic systems, establishing relations between their properties or state variables, such as their internal energy $U$, temperature $T$, entropy $S$, pressure $P$, and volume $V$. For infinitesimal processes connecting infinitely near equilibrium states of a closed system, i.e. a system that exchanges no matter with its surroundings [1], these variables are related by the *system fundamental equation* [1–3]

$$\mathrm{d}U = T\mathrm{d}S - P\mathrm{d}V \,. \tag{1}$$

These infinitesimal processes are regarded as the building blocks of any finite process. Our study is restricted to closed systems, but for an open system the term $\mu\mathrm{d}N$ would be added to the RHS of (1), where $\mu$ is the chemical potential and $N$ is the number of moles of the system [1].

Temperature and entropy are universal variables of all systems in thermodynamic equilibrium. Pressure and volume, on the other hand, are not useful for characterising some systems; in these cases, $P$ and $V$ are replaced by more appropriate mechanical variables, called *generalised forces* and *generalised displacements*, respectively [1]. As an example, suitable variables for an electric battery are the electromotive force, which plays the role of $P$, and the electric charge, which plays the role of $V$. However, without losing generality, we will proceed by considering $P$ and $V$ in (1), because these variables are more familiar and intuitive.

A thermodynamic process is an interaction between the system and surroundings, the latter characterised by the counterpart variables of those in (1) and satisfying a formally identical equation. So, using the subscript 'e' to denote surroundings variables, we obtain

$$\mathrm{d}U_\mathrm{e} = T_\mathrm{e}\mathrm{d}S_\mathrm{e} - P_\mathrm{e}\mathrm{d}V_\mathrm{e} \,. \tag{2}$$

Regarding an interaction, the laws of physics do not depend on which is labelled as the system or as the surroundings. So, using the *principle of conservation of energy*, we have

$$\mathrm{d}U = -\mathrm{d}U_\mathrm{e} \,, \tag{3}$$

which, in conjunction with (1) and (2), leads to the *process fundamental equation*,

$$T\mathrm{d}S - P\mathrm{d}V = -T_\mathrm{e}\mathrm{d}S_\mathrm{e} + P_\mathrm{e}\mathrm{d}V_\mathrm{e} \,. \tag{4}$$

Equation (4) is invariant under a system–surroundings interchange. This formally implies that when the subscript 'e' is removed from the surroundings variables and assigned to the system variables, an indistinguishable equation is obtained. This property is referred to as *process invariance* [4]. Two other relevant process invariants, both non-negative valued, are the *dissipative work* [5],

$$\delta W_\mathrm{D} = P\mathrm{d}V + P_\mathrm{e}\mathrm{d}V_\mathrm{e} = T\mathrm{d}S + T_\mathrm{e}\mathrm{d}S_\mathrm{e} \geq 0 \,, \tag{5}$$

and the *entropy generation*,

$$\mathrm{d}S_\mathrm{G} = \mathrm{d}S + \mathrm{d}S_\mathrm{e} \geq 0\,. \tag{6}$$

Incorporating system and surroundings variables, as well as the principle of conservation of energy, equation (4) is powerful because it contains *all the thermodynamic information about the process* [4]. Equation (5) defines the *dissipative work.* As we will see in the next section, when divided by the temperature of the system, this is one of the contributions to the entropy generation; it must therefore have a non-negative value. Equation (6) represents *entropy generation* and states the *second law of thermodynamics,* providing the ultimate criterion for evaluating whether the process is reversible (in the equality case) or irreversible (in the greater-than case). These equations, being process invariants, constitute the cornerstones of the theory and are therefore a guide to the establishment of other concepts, which, while useful, cannot add anything fundamentally new.

The concepts of *heat* and *work*, although basic, do not naturally follow from (4). The same is true for the concepts of *lost work* and *exergy*. These concepts – *heat*, *work*, *lost work*, and *exergy* – require additional assumptions that must be clearly stated and justified to dispel disagreements or ambiguities. This is the subject of the remainder of this paper.

## 2. Irreversibility in the system and lost work

In contrast to a process invariant, a *flow* changes sign under a system–surroundings interchange and can therefore be understood as a physical quantity that crosses the system boundary. The principle of energy conservation, expressed by (3), allows us to understand the internal energy variation $\mathrm{d}U$ as the result of an energy flow.

Unlike energy, entropy $S$ is not conserved in irreversible processes. This non-conservation is perhaps the feature that most hinders the understanding of entropy and related concepts such as *heat* and *lost work.* Therefore, neither $\mathrm{d}S$ nor $\mathrm{d}S_\mathrm{e}$ are flows; that is, in general, $\mathrm{d}S \neq -\mathrm{d}S_\mathrm{e}$. However, these entropy variations can be expressed in terms of the *entropy flow* $\mathrm{d}S_\phi$ as

$$\mathrm{d}S = \mathrm{d}S_\phi + \beta\,\mathrm{d}S_\mathrm{G}\,, \tag{7}$$

$$\mathrm{d}S_\mathrm{e} = -\mathrm{d}S_\phi + \beta_\mathrm{e}\,\mathrm{d}S_\mathrm{G}\,, \tag{8}$$

where $\beta$ and $\beta_\mathrm{e}$ are two *arbitrary* non-negative parameters indicating the fractions of entropy generation occurring in the system and in surroundings, respectively, thus satisfying

$$\beta + \beta_\mathrm{e} = 1\,. \tag{9}$$

What has physical meaning is the entropy generation $\mathrm{d}S_\mathrm{G}$ as a whole, and not where it comes from. For a given infinitesimal process, $\mathrm{d}S$ and $\mathrm{d}S_\mathrm{e}$ are uniquely determined, and, by (6), so is $\mathrm{d}S_\mathrm{G}$. Besides, (4)–(6) are independent of $\beta$, $\beta_\mathrm{e}$, and $\mathrm{d}S_\phi$, thus these values are arbitrary, the only restriction being that they must satisfy (7)–(9). In other words, all processes that differ *only* by the values of $\beta$, $\beta_\mathrm{e}$, and $\mathrm{d}S_\phi$ are identical [6], i.e., indistinguishable. So, the entropy flow $\mathrm{d}S_\phi$ is not defined for a given process unless specific assumptions are made.

Of all the identical processes corresponding to different values of $\beta$ and $\beta_\mathrm{e}$, we then choose a *gauging process* for which $\mathrm{d}S_\phi$ has a definite and unique value. In this process, entropy generation occurs solely in the system; consequently, the surroundings are said to consist of *reservoirs*, the defining characteristic of which is the absence of entropy generation within them [4].

Therefore, the reservoir concept determines the gauging process, defined by the conditions

$$\beta = 1, \tag{10}$$

$$\beta_\mathrm{e} = 0, \tag{11}$$

which turn (7) and (8) into

$$\mathrm{d}S = \mathrm{d}S_\phi + \mathrm{d}S_\mathrm{G}, \tag{12}$$

$$\mathrm{d}S_\mathrm{e} = -\mathrm{d}S_\phi. \tag{13}$$

Inserting (12) and (13) into (4), the latter becomes

$$T\mathrm{d}S_\phi + T\mathrm{d}S_\mathrm{G} = T_\mathrm{e}\mathrm{d}S_\phi + \delta W_\mathrm{D}. \tag{14}$$

This equation is equivalent to (4) but incorporates the *reservoir concept* while relating two process invariants, $\mathrm{d}S_\mathrm{G}$ and $\delta W_\mathrm{D}$, and a flow, $\mathrm{d}S_\phi$. The $\delta W_\mathrm{D}$ term is the *dissipative work*, as set out in (5).

The $T_\mathrm{e}\mathrm{d}S_\phi$ term in (14) is the *heat* $\delta Q$ [4]; thus, from (13), we have

$$\delta Q = T_\mathrm{e}\mathrm{d}S_\phi = -T_\mathrm{e}\mathrm{d}S_\mathrm{e}. \tag{15}$$

Once the heat $\delta Q$ has been defined, the work $\delta W$ is obtained using the *first law of thermodynamics*. This law stated as [1]

$$\mathrm{d}U = \delta Q + \delta W, \tag{16}$$

together with (2) and (3), leads to

$$\delta W = P_\mathrm{e}\mathrm{d}V_\mathrm{e}. \tag{17}$$

At this point, it is worth noting that the signs of $\mathrm{d}S_\phi$, $\delta Q$ and $\delta W$ follow naturally from (12) and (16). They are, therefore, positive if entropy or energy enters the system, leading to an increase in *S* or *U*, respectively. Conversely, they are negative if entropy or energy leaves the system, leading to a decrease in *S* or *U*, respectively.

The $T\mathrm{d}S_{\mathrm{G}}$ term in (14) is the *lost work* $\delta W_{\mathrm{L}}$ [4],

$$\delta W_{\mathrm{L}} = T\mathrm{d}S_{\mathrm{G}}\,, \tag{18}$$

and using (13)–(15), it can be written as

$$\delta W_{\mathrm{L}} = \left(1 - \frac{T}{T_{\mathrm{e}}}\right)\delta Q + \delta W_{\mathrm{D}}\,. \tag{19}$$

To understand why $\delta W_{\mathrm{L}}$ is referred to as lost work, let us look at Figure 1, specifically, the dashed lines. The first term in the RHS of (19) is the *lost work due to thermal irreversibility*: the work we would obtain from a Carnot engine (CE) operating between the temperatures $T$ and $T_{\mathrm{e}}$. The second term in the RHS of (19) is the *lost work due to mechanical irreversibility*, which is the direct conversion of work into internal energy.

The system–surroundings interaction is shown in Figure 1, which, in addition to the terms on the RHS of (19), shows the heat $\delta Q$ and work $\delta W$, as given by (15) and (17), respectively. Irreversibility occurs only in the system because the surroundings consist of reservoirs.

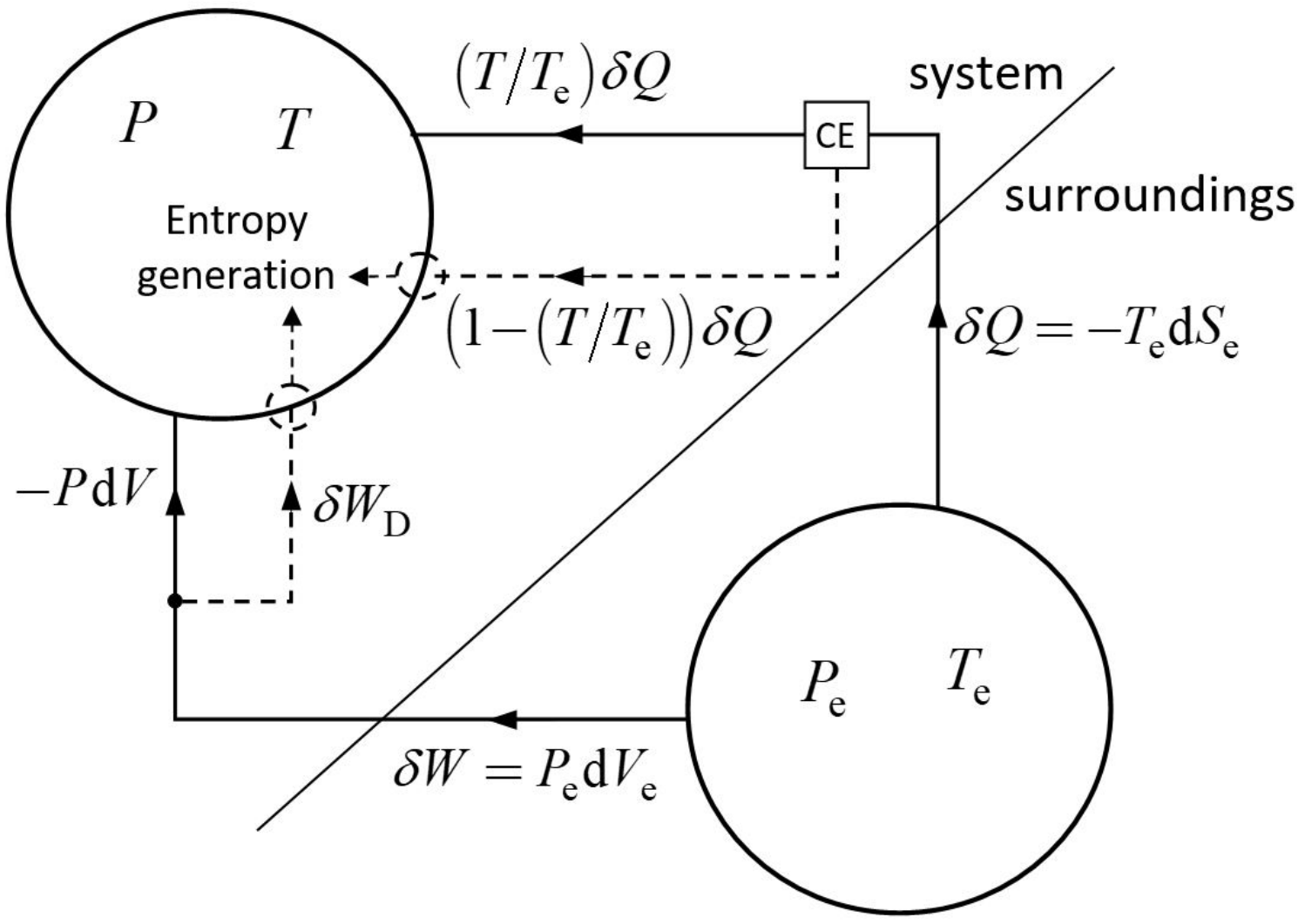

**Figure 1.** The thermodynamic process is a system–surroundings interaction, in which the surroundings consist of reservoirs. The heat $\delta Q$, work $\delta W$, lost work $\delta W_{\mathrm{L}}$ (dashed lines), and entropy generation, which occurs only in the system, are represented.

The reservoir concept imposes an asymmetry on the system–surroundings interaction, which is a drawback of having unique definitions of heat, work, and entropy flow. This is consistent with Clausius' relation [7]; indeed, as $\delta Q = -T_{\mathrm{e}}\mathrm{d}S_{\mathrm{e}}$, for a cyclic process, that is, one in which the surroundings drive the system back to its initial state, we have

$$\oint \left(\mathrm{d}S + \mathrm{d}S_{\mathrm{e}}\right) = \oint \mathrm{d}S_{\mathrm{e}} = \oint \frac{-\delta Q}{T_{\mathrm{e}}} \geq 0\,. \tag{20}$$

However, the description of a thermodynamic process given so far, illustrated in Figure 1, must be modified in engineering contexts where the system is said to be ‘internally reversible’, that is, instead of being in the system, the irreversibility is in the surroundings. This is the case when one compares a given irreversible process with a reversible process that takes the system along the same path. How would the model in Figure 1 need to be adjusted if irreversibility occurred in the surroundings? We answer this question in the following section.

## 3. Irreversibility in the surroundings and exergy

To model an internally reversible system, we redirect the dashed lines in Figure 1, which represent the lost work, towards the surroundings, as shown in Figure 2. The new equations that describe the process are then obtained by interchanging the system and surroundings variables in the original equations. The entropy generation $\mathrm{d}S_{\mathrm{G}}$ and dissipative work $\delta W_{\mathrm{D}}$ remain unchanged because they are process invariants. Thus, using the prime symbol to denote the quantities after interchanging variables, (15), (17), and (19) become

$$\delta Q' = -T\mathrm{d}S\,, \tag{21}$$

$$\delta W' = P\mathrm{d}V\,, \tag{22}$$

$$\delta W_{\mathrm{L}}' = \left(1 - \frac{T_{\mathrm{e}}}{T}\right)\delta Q' + \delta W_{\mathrm{D}}\,. \tag{23}$$

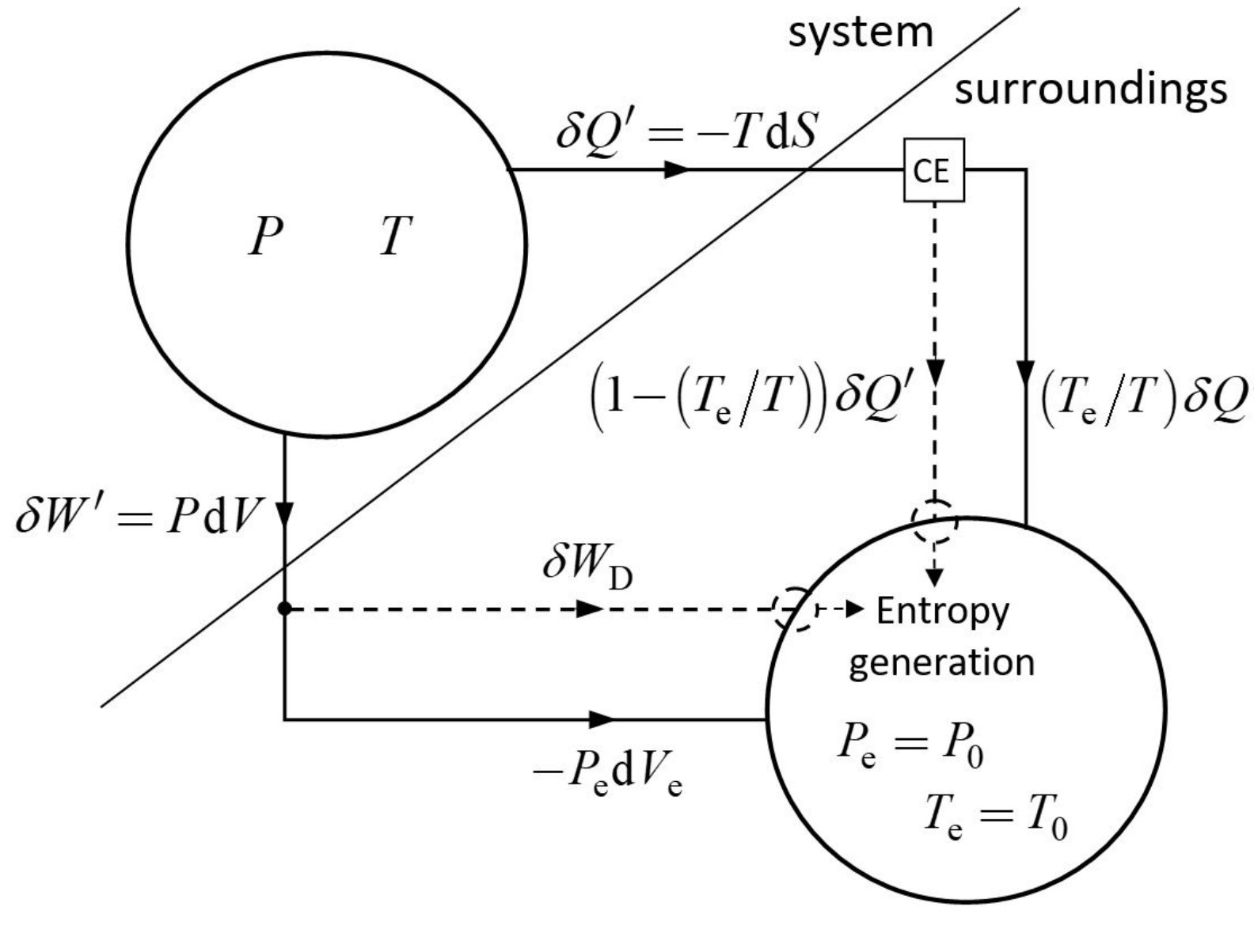

**Figure 2.** The same process as in Figure 1, but now with irreversibility occurring in the surroundings. The new values of the heat $\delta Q'$, work $\delta W'$, lost work $\delta W_{\mathrm{L}}'$ (dashed lines), and entropy generation, which now occurs only in the surroundings, are shown. In this case, the lost work is called *exergy*, which is a system property, while the surroundings are maintained at a constant temperature $T_0$ and a constant pressure $P_0$.

Since the lost work is positive and now goes to the surroundings, heat and work are now positive when they leave the system because we have kept the system the same as before (in fact, it now plays the same role as the surroundings did before). However, the signs are not relevant because physics does not depend on what we have labelled as the system or as the surroundings.

In engineering, in addition to often being considered to contain the irreversibility, the surroundings are commonly taken to be at constant temperature $T_0$ and constant pressure $P_0$, that is, $T_\mathrm{e} = T_0$ and $P_\mathrm{e} = P_0$, as shown in Figure 2. Thus, using (1), (5), and (21), and considering $\mathrm{d}V = -\mathrm{d}V_\mathrm{e}$, the lost work given by (23) becomes

$$\delta W'_\mathrm{L} = -\mathrm{d}U + T_0 \mathrm{d}S - P_0 \mathrm{d}V \,. \tag{24}$$

If the system is brought into thermodynamic equilibrium with the surroundings, integrating (24) yields

$$W'_\mathrm{L} = \left(U - U_0\right) - T_0\left(S - S_0\right) + P_0\left(V - V_0\right), \tag{25}$$

where $U_0$, $S_0$, and $V_0$ are values for the system upon reaching equilibrium. For surroundings at constant temperature $T_0$ and at constant pressure $P_0$, $W'_\mathrm{L}\left(U,S,V\right)$ is a state function, or system property, referred to as *exergy* [8–10]. The exergy is positive, except when the system is in equilibrium with the surroundings, in which case it is zero, that is, $W'_\mathrm{L}\left(U_0,S_0,V_0\right) = 0$.

In conjunction with (5), (6), and (21), (18) and (23) yield

$$\mathrm{d}S_\mathrm{G} = \frac{\delta W_\mathrm{L}}{T} = \frac{\delta W'_\mathrm{L}}{T_0} \geq 0 \,, \tag{26}$$

and therefore $\delta W_\mathrm{L}$ and $\delta W'_\mathrm{L}$ are both closely related to the fundamental concept of entropy generation. From (26), we obtain

$$\frac{\delta W_\mathrm{L}}{\delta W'_\mathrm{L}} = \frac{T}{T_0} \,, \tag{27}$$

and when $T = T_0$, that is, in the absence of thermal irreversibility, the two concepts coincide, and, by (19) and (23), are both equal to $\delta W_\mathrm{D}$.

It is instructive to emphasize that Figures 1 and 2 show two alternative descriptions of a given thermodynamic process. The values of $\mathrm{d}U$, $\mathrm{d}S$, $\mathrm{d}U_\mathrm{e}$, and $\mathrm{d}S_\mathrm{e}$ are the same in each case and, consequently, the entropy generation is also the same. Once the process is finished, it is not possible to determine which of the descriptions has been adopted and, in this sense, the two are indistinguishable [6]. Despite being the same process, the heat, work, lost work, and entropy flow have different values for the two descriptions. This is possible because these quantities are not state functions and cease to exist once the process is complete.

## 4. The physical meaning of exergy

Figure 3 provides a physical interpretation of exergy. If the lost work $\delta W_{\mathrm{L}}'$ in Figure 2 were entirely recovered as useful work, through a *work reservoir* [11] placed in the surroundings, the process would become reversible, although the system would still follow the same path as in the irreversible process. Therefore, exergy is the maximum useful work that can be obtained from a process in which the system reversibly reaches equilibrium with its surroundings [9].

It is worth asking whether we can obtain the same useful work from $\delta W_{\mathrm{L}}$ in Figure 1. The answer is yes, but in this case, the system would not reach the same final state as in the irreversible process. When we replace a given irreversible process with a reversible one, for which the system evolves between the same states, the final state of the surroundings cannot be the same as in the irreversible process. Conversely, if the surroundings evolve between the same states in the reversible process as in the irreversible process, the system does not.

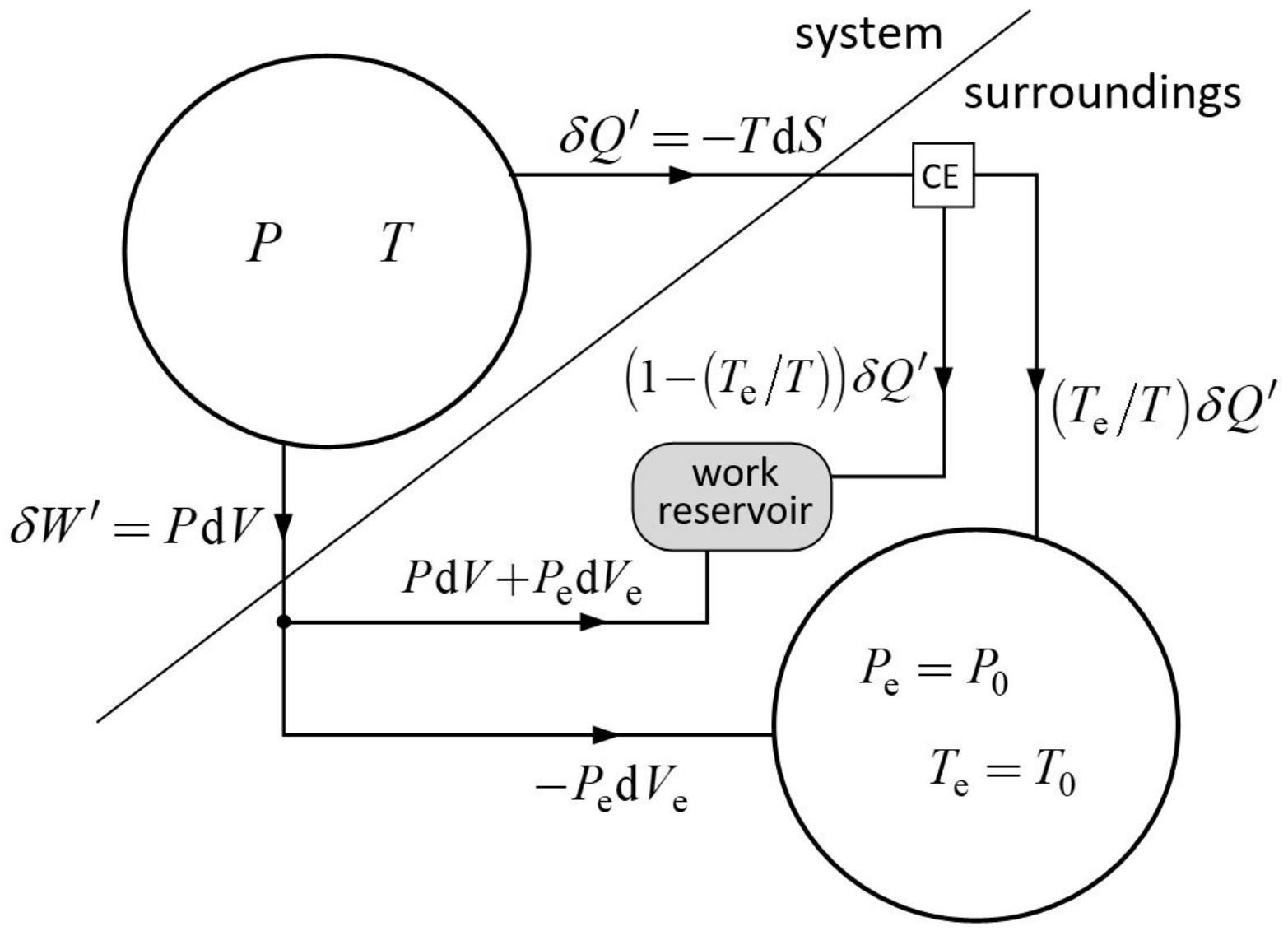

**Figure 3.** An illustration of the physical meaning of exergy. The lost work in Figure 2 is recovered as *useful work* so that the process becomes reversible, but the system follows the same path until it reaches equilibrium with its surroundings. This useful work is the exergy of the system in its initial state.

## 5. Conclusions

As discussed, for a closed system, an infinitesimal process seen as a building block of a finite process is fully described by the fundamental equation (4). Even though this equation contains all the thermodynamic information about the process, to set up the concepts of *heat*, *work*, and *lost work*, it is also necessary to specify where irreversibility occurs, because this location is intrinsically indeterminate.

The usual assumption in physics is to take the surroundings as consisting of reservoirs, whose defining property is the absence of irreversibility, or, equivalently, the non-existence of entropy generation within them. Therefore, the irreversibility is confined to the system, which corresponds to (10) and (11). The concept of reservoir, thus introduced, is indispensable for unequivocal definitions of heat, work, and lost work through (15), (17) and (19), respectively. In addition, the Clausius relation (20), which is an important milestone in classical physics, is readily established, which shows the strength of this formalism.

However, engineering typically has a different theoretical perspective, which leads to other difficulties. Conceptual confusion can arise in the connection between physics and engineering and, to the best of our knowledge, this is the first study to clarify this matter. In contrast to physics, engineering considers irreversibility in the surroundings, which implies that the definitions of heat, work, and lost work are changed to (21), (22) and (23), respectively, while the lost work becomes what is known in the literature as exergy.

Therefore, *lost work* and *exergy* are ultimately different views of the same concept, which in turn is closely related to entropy generation. While *lost work* stems from irreversibility being considered in the system, *exergy* is the lost work when irreversibility is considered in the surroundings instead. The physical meaning of exergy was also discussed and stated to be the maximum useful work that can be obtained from a process in which the system reversibly reaches equilibrium with its surroundings. The discussion in this study, as well as the diagrams used in the argumentation, is novel and instructively relevant, as it promotes a global and consistent perception of the important concepts, thus avoiding misunderstandings that are often encountered in thermodynamics. Furthermore, it provides a clarification of the conceptual differences between physics and engineering approaches, thereby making it useful not only for educators, but also for researchers in conceptual thermodynamics.